\documentclass[aps,onecolumn,amsmath,amssymb,groupedaddress,notitlepage,floatfix]{revtex4-2}

\usepackage[colorlinks=true, linkcolor=Blue, allcolors=Blue]{hyperref}
\usepackage{graphicx}
\usepackage[usenames, dvipsnames]{xcolor}
\usepackage{fancyhdr} 
\usepackage{newtxtext,newtxmath}
\usepackage{dcolumn}
\usepackage{bm}
\usepackage{subfig}
\usepackage[english]{babel}

\begin{document}

\title{Jet creation and wave energy dissipation by a submerged elastic plate}
\author{Diane Komaroff\textsuperscript{1}}
\author{Gatien Polly\textsuperscript{2}}
\author{Alexis M\'erigaud\textsuperscript{3}}
\author{Benjamin Thiria\textsuperscript{1}}
\author{Ramiro Godoy-Diana\textsuperscript{1}}
\affiliation{\textsuperscript{1}Laboratoire de Physique et M\'ecanique des Milieux H\'et\'erog\`enes (PMMH), CNRS UMR 7636, 75005 Paris, France}
\affiliation{\textsuperscript{2}IFP \'Energies Nouvelles, 92852 Rueil-Malmaison, France}
\affiliation{\textsuperscript{3}Saint-Gobain Research Paris, 93300 Aubervilliers, France}

\begin{abstract}
Experiments on a flexible plate horizontally submerged in a wave tank are reported in this paper. The plate is held at a fixed depth at its upstream end, while its downstream end is free to move. The wave field is measured using a synthetic Schlieren technique, giving access to an estimate of the fractions of the incoming wave energy that are transmitted and reflected after encountering the region where the plate is submerged. Part of the incoming wave energy is shown to be dissipated  by the plate. Flow field measurements in a vertical plane parallel to the wave propagation are performed to examine the wave-plate interaction. A jet-like flow is produced at the free end of the plate, with strong vortex activity similar to that produced by a flapping fin. The kinetic energy  present in this jet-like flow is computed from the velocity field measurements, and the external dissipation produced by jet creation is estimated. Results highlight the efficiency of flexible plates in reflecting and attenuating waves. This study also showcases a change in jet direction, in particular the observation of a jet going upward toward the free surface for certain frequencies. Such a configuration, could be of great interest for wave energy conversion applications, while preventing harmful phenomena such as seabed erosion.

\end{abstract}


\maketitle
\thispagestyle{fancy}

\section{Introduction}
Waves are a promising renewable energy source \cite{Mork2010,Kempener2014}. Despite the abundance of preliminary designs of wave energy harvesting technologies, fundamental questions surrounding wave energy conversion scenarios remain elusive. Practical applications are now often discussed intermingled with questions of coastal management and protection \cite{Mustapa2017}, where a single system or a farm are examined with a double goal in mind, energy harvesting, and wave attenuation for coastal protection.

Flexible body wave energy converters (WECs) have considerable potential to withstand the harsh conditions endured by marine energy harvesting devices better than rigid WECs \cite{Collins2021}. In that context, flexible membranes have been at the centre of recent research efforts, in particular following the development of new dielectric elastomer materials \cite{Zheng2022,Gurjar2025, Wang2025}. 

Research on submerged flexible plates has predominantly examined two-edge-fixed configurations for energy extraction and coastal protection. Early studies by Cho and Kim \cite{Cho1998, Cho2000} investigated these plates as wave barriers, while Alam \cite{Alam2012} proposed a single-hinged flexible plate system mimicking natural seabed damping mechanisms. This "Wave-Carpet" concept has been explored through numerical and experimental studies \citep{Desmars2018, Asaeian2020, Lehmann2013}, with extensions incorporating piezoelectric materials \citep{Renzi2016}, wave-current interactions \citep{Achour2020}, and elastic foundation modelling \citep{Boral2023}.
Complex two-edge-clamped configurations have also been investigated, including plates near vertical structures \citep{Guo2020, Gayathri2020}, multi-layer arrangements \citep{Mohapatra2014, Mohapatra2019, Das2020}, and sequential arrays \citep{Mirza20241, Mirza20242}.
Single-edge configurations have received limited attention. Shoele \cite{Shoele2023} studied hybrid wave-current converters numerically without isolating wave effects, while Michele et al. \cite{Michele2020} examined floating elastic plates with multiple PTO units, demonstrating that flexibility introduces beneficial resonant frequencies that expand operational bandwidth beyond rigid designs. 

A recent small-scale study by Polly et al. \cite{Polly2025} examined the wave-structure interaction of a flexible submerged elastic plate, placed horizontally and clamped at one edge. Transmission and reflection coefficients for a range of wavelengths and amplitudes have been investigated, and regimes that may offer enhanced performance in applications such as coastal protection, when compared with a rigid plate, are identified. The present study follows up on \cite{Polly2025} using a similar experimental setup but in the case of a more rigid plate to avoid it going up to the free surface. PIV experiments are performed to examine the fluid dynamical mechanisms behind the strong dissipation of incoming waves by the submerged plate. In particular, for certain wave frequencies, it is observed that a significant fraction of the wave energy is dissipated by vortices shed at the edges of the plate, giving rise to a jet similar to that observed for a flapping fin. We quantify how incoming wave energy interacts with the flexible plate using two measurement approaches: 2D wave field measurements throughout the experimental tank and bulk velocity field measurements around the submerged plate. The values of the transmission and reflection coefficients demonstrate how flexibility promotes dissipation. Results also show that, unlike experiments conducted with fixed rigid plates, vortices are not advected and remain close to the tip of the plate. Finally, as frequencies increase, the jet changes direction: whilst at high frequencies it is directed towards the bottom of the tank, at lower frequencies the jet goes towards the free surface, which has not been previously reported in other studies of flexible plates forced by waves.

\newpage
\section{Experimental setup and methods}
\subsection{Wave tank and plate}
\begin{figure}[t]
\centering
\subfloat[]{\includegraphics[scale=0.45]{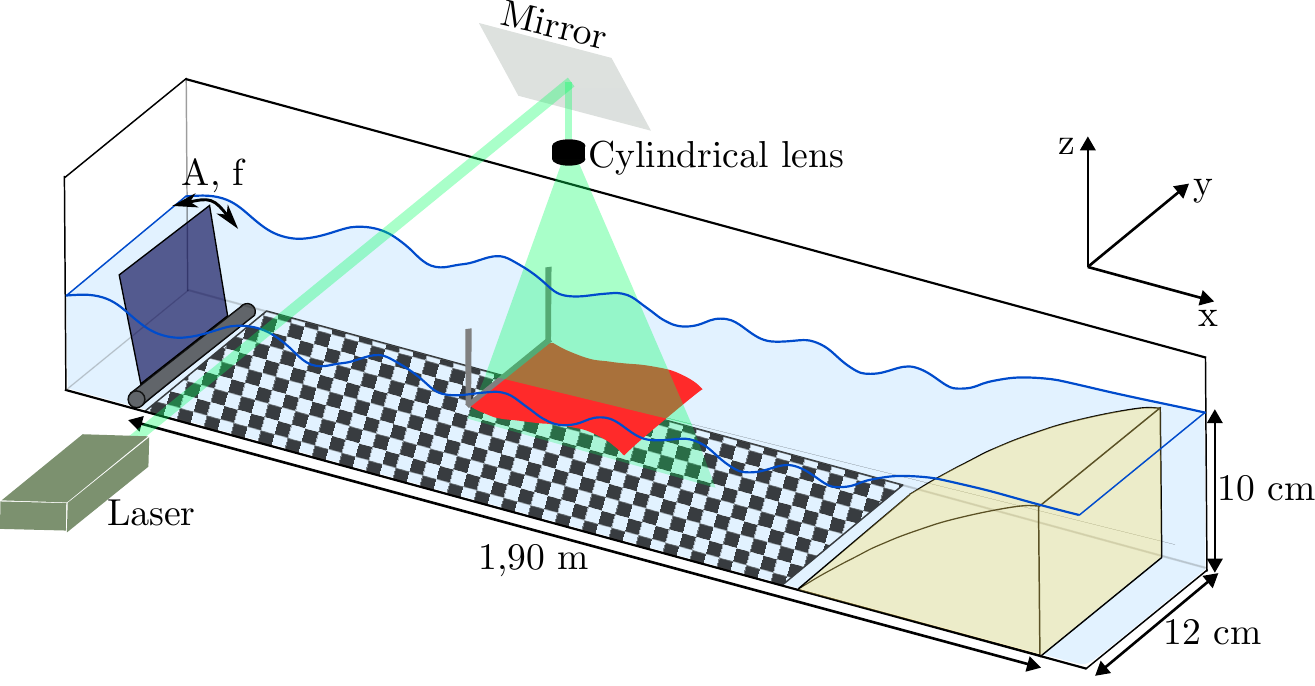}\label{Fig: Method}}\hspace{1cm}
\subfloat[]{\includegraphics[scale=0.55]{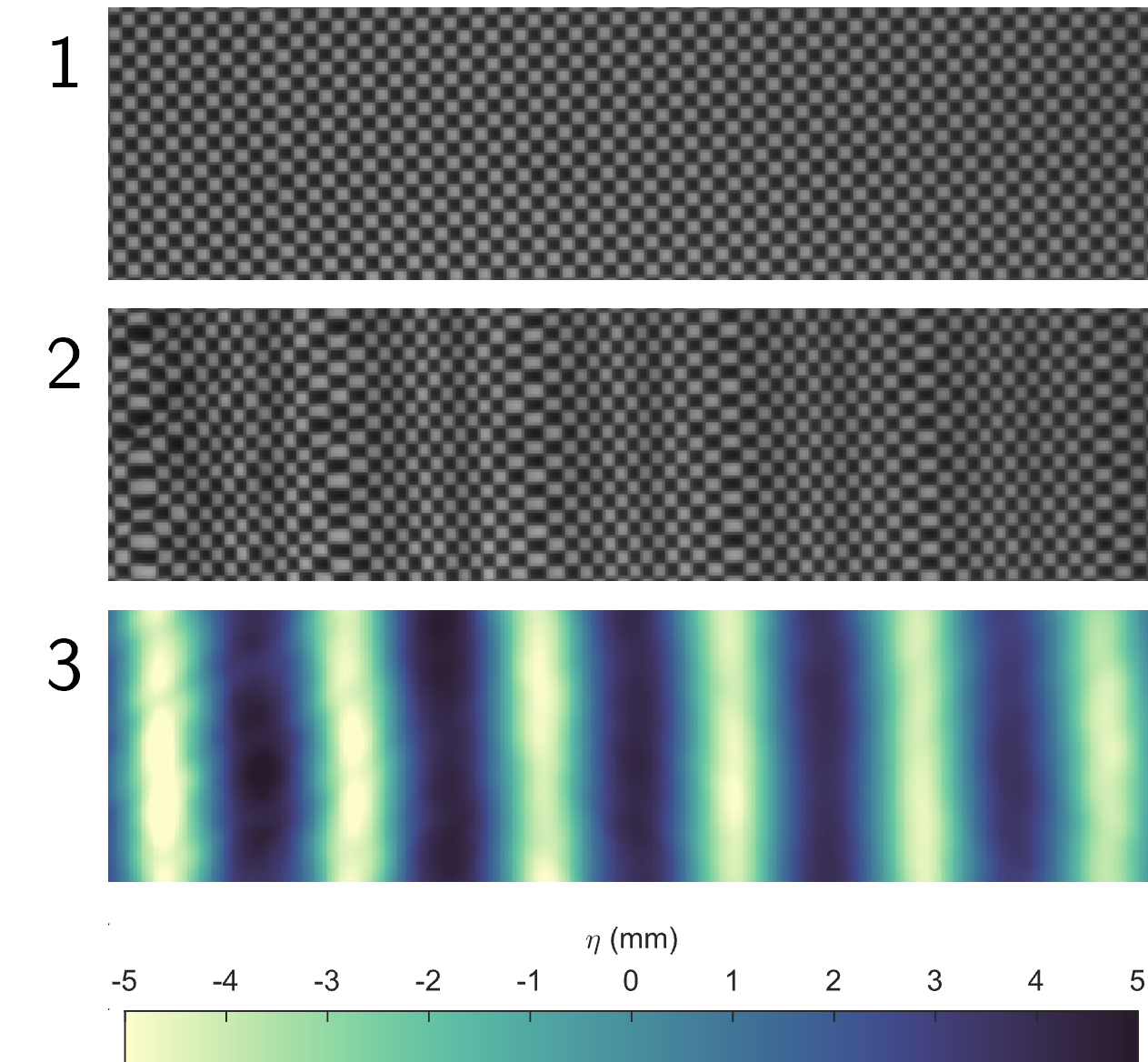}\label{Fig: Schlieren}}
\caption{(a) Scheme of the experimental set-up. The plate length is $L=19$ cm.
(b) Example of Schlieren visualisation. 1: Top view of the tank when no waves are propagating, the checkerboard is not deformed. 2: Top view of the tank when waves of $f=3$ Hz and $A=4$ mm amplitude are propagating. 3: Surface height calculated with Schlieren treatment.}
\end{figure}

Experiments are carried out inside a wave tank illustrated in Figure~\ref{Fig: Method}. $x$ indicates the direction of wave propagation, $y$ represents the transverse direction, and $z$ denotes the vertical axis. The tank is 2.5 metres long, with a central measurement region spanning 2 metres. In order to restrict transverse modes, the tank width is fixed at $b=12$ cm \cite{Ursell1952}. The water depth $h$ remains constant at 10 centimetres.
Located at the midpoint of the length of the tank is a plate, while a parabolic beach is placed at the end of the tank to minimise reflections from the tank wall, following the methodology outlined by \cite{Ouellet1986}. The beach features 5-millimeter circular perforations to increase viscous damping.
The plate is submerged at a fixed depth of 3 cm. It is made of polycarbonate of density $\rho_p=$ 1200 kg.m$^{-3}$ and stiffness $EI=$ 7.6.10$^{-3}$ N.m$^2$. The material is chosen in order to have a small density difference with water and thus minimise deviation from the horizontal at rest. The plate length is $L=$ 19 cm, its width is $l=$ $10$ cm, and its thickness is $e=750 \mu$m. The leading edge of the plate is maintained with a clamping system composed of two carbon rods of 2 mm diameter glued to the edge of the plate. These rods are attached to the support poles on the sides of the tank, ensuring that the leading edge remains still. 
The plate resonant frequencies are obtained by performing a forced vibration test of the submerged plate: the natural resonant frequency is found to be $f_0=0.68$ Hz.

A flap-type wave-maker, driven by a Linmot\textsuperscript{\textregistered} linear motor model DM01-23x80F-HP-R-60\textunderscore MS11, is used to generate the waves. Wave amplitudes $A$ range from 0.5 to 5 millimeters and frequencies $f$ vary between 1.5 and 4 Hz. The relationship between the wave's angular frequency, $\omega$, and the wavenumber, $k$, is governed by the gravity wave dispersion relation:
\begin{equation}
\omega^2=gk\tanh(kh),
\end{equation}
with $g$ acceleration of gravity and $h$ water depth. The investigation encompasses wavelengths, $\lambda$, ranging from 10 to 55 centimetres, corresponding to $\lambda/h$ ratios ranging from 1 to 5.5, indicating deep to intermediate water depths.
Data acquisition from this setup involves either filming from the top or side perspectives of the tank, enabling observation of free surface deformation, plate motion, and illuminated particles for velocimetry measurements.

\subsection{Particle image velocimetry setup and processing}\label{section PIV}

\begin{figure}[t]
\centering
\includegraphics[width=\linewidth]{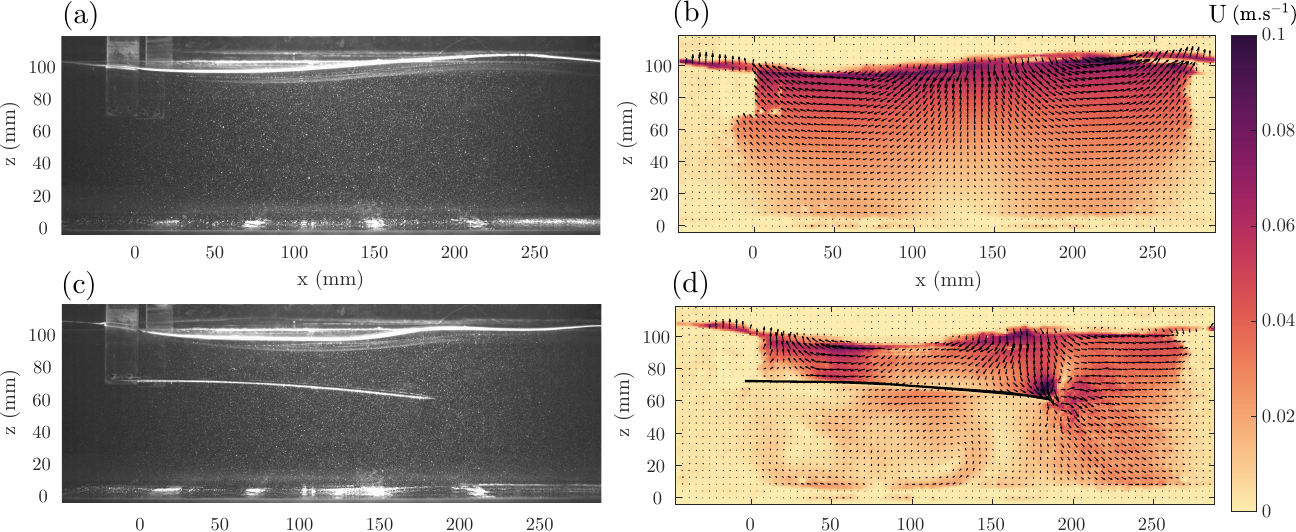}
\caption{(a) and (c) Instantaneous particle images, with (b) and (d) their respective instantaneous velocity fields (phase averaged, see text), for the case without (a, c) and with (b, d) the plate. The case presented is for waves of amplitude $A=4$ mm and frequency $f=2.1$ Hz. The field shown in color is the velocity norm $U=\sqrt{U_x^2+U_z^2}$.}
\label{Fig:PIV_with_and_without_plate}
\end{figure}

Two-dimensional particle image velocimetry (PIV) is employed to observe fluid motion. The fluid is seeded with particles, then, in order to observe the flow velocity in a two-dimensional plane, a laser sheet in the $x-z$ plane illuminates the particles as represented in Figure~\ref{Fig: Method}.  By rotating the light sheet optics, the laser sheet can also be placed in the $y-z$ plane. Particles located in the laser sheet diffract laser light and appear brighter. Polyamide seeding particles of 20 $\mu$m diameter and 1.03 density are used. The choice was made not to use larger particles as passive tracers to mitigate particle sedimentation, as the waves do not generate a significant mean flow. 

A continuous 532 nm laser operating at 1 W was used. A 1 mm wide laser sheet is formed and directed using a cylindrical lens and a mirror, resulting in the configuration presented in Figure~\ref{Fig: Method}.
A side-view camera records particles and plate motion. The side-view frame rate is adjusted based on the wave frequency using the relation $f_\mathrm{acq}$ = 40$\times f$, where $f$ represents the wave frequency and $f_\mathrm{acq}$ signifies the frame rate. This approach ensures sufficient image data for phase averaging over each period, with the exposure time set at 1/$f_\mathrm{acq}$. Velocity fields are subsequently computed using the Davis software developed by LaVision\textsuperscript{\textregistered}. For the experiments, correlation windows of $12\times 12$ pixels were used, the observed flows produced maximum displacement of 5 pixels. The physical area of the measurement window is (100 x 250 mm) see Figure \ref{Fig:PIV_with_and_without_plate}. Fig. \ref{Fig:PIV_with_and_without_plate} (a) and (c) show instantaneous image of the seeded fluid when waves of $A=4$ mm and $f=2.1$ Hz are propagating, in the case where the plate is absent (a) and present (c), showing how the waves deform when interacting with the plate. The corresponding instantaneous velocity fields are represented in Fig. \ref{Fig:PIV_with_and_without_plate} (b) and (d). Clear influence of the plate is observed with almost no more fluid displacement in the region under the plate and the apparition of high and oriented velocity at the tip of the plate. It can be seen that the laser sheet does not cover the full field of view of the camera, so no vector field can be computed before $x=0$ mm and the quality of the PIV calculation degrades for $x\gtrsim 260$ mm.

\subsection{Wave height measurements}

Free surface deformation is obtained using Schlieren Imaging \cite{Moisy2009,Wildeman2018}. A checkerboard pattern is placed under the tank and filmed with a top view camera. An example of the pattern deformation and free surface elevation after treatment is illustrated in Figure \ref{Fig: Schlieren}. Free surface reconstruction is done using the open-source code provided in \cite{Wildeman2018}. The deformation, $\mathbf{u}$, measured numerically is linked to the free surface deformation, $\eta$, by the relation  \cite{Moisy2009}:
\begin{equation}
    \nabla\eta=-\frac{\mathbf{u}}{h^*},
\end{equation}
with $h^*$ a coefficient depending on water depth and setup conditions:
\begin{equation}
    h^*=\left(1-\frac{n_a}{n_w}\right)\left(h+\frac{n_w}{n_g}h_g\right),
\end{equation}
$n_a$, $n_w$, and $n_g$ being the optical index of air, water and glass respectively, and $h_g$ the tank glass thickness.

Following the work in \cite{Polly2025} we divide the tank into two regions. An up-wave region located between the wave-maker and the plate, and a down-wave region located between the plate and the beach. In the up-wave region we have the incoming wave produced by the wave-maker and the wave reflected by the plate. In the down-wave region we only consider transmitted waves as almost no waves are reflected from the absorption beach (less than 10\% for waves of amplitude $A$ close to  4 mm, which corresponds to our experimental condition \cite{Polly2023_PhDthesis}). Before the object, the elevation of the free surface can be written as the sum of two waves, one propagating forwards and the other propagating backwards:
\begin{equation}\label{eq etai+etar}
    \eta^{\text{uw}}(x,t) = \text{Re}\left(\underline{\eta} _i e^{-i(\omega t - kx)-\nu x} +\underline{\eta}_r e^{-i(\omega t + kx)+\nu x} \right),
\end{equation}
where $\underline{\eta}_i$ is the complex-valued amplitude of the incident wave propagating forward, $\underline{\eta}_r$ the complex-valued amplitude of the reflected wave propagating backward and $\nu$ the damping coefficient modelling wave dissipation along the tank. We define the amplitude reflection coefficient, $K_r$ as:
\begin{equation}\label{Kr}
    K_r = \left( \frac{|\underline{\eta_r}|}{|\underline{\eta_i}|} \right)^2 .
\end{equation}
After the object, only the transmitted wave propagates and the elevation of the free surface is written : 
\begin{equation}
    \eta^{\text{dw}}(x,t) = \text{Re} \left(\underline{\eta}_t e^{-i(\omega t - kx)-\nu x} \right) ,
\end{equation}
where $\eta_t$ is the amplitude of the transmitted wave. 
The amplitude transmission coefficient, $K_t$, is defined as:
\begin{equation}\label{Kt}
    K_t = \left( \frac{|\underline{\eta_t}|}{|\underline{\eta_i}|} \right)^2.
\end{equation}

\section{Results}

\subsection{Energy reflection and transmission coefficients}

\begin{figure}[t]
\centering
\includegraphics[width=\linewidth]{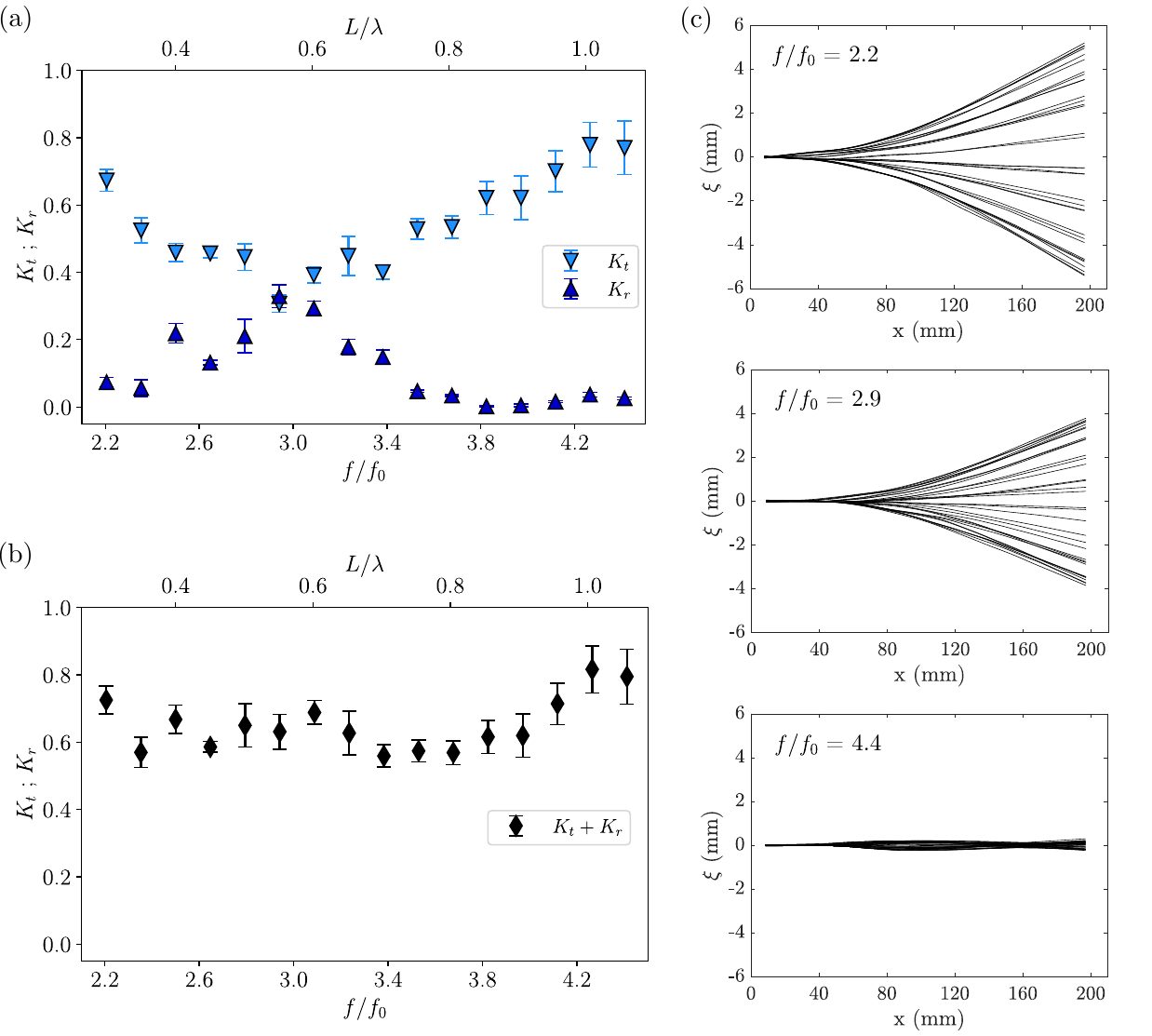}
\caption{(a) Average over five experiments of the reflection and transmission coefficients, $K_r$ (triangles) and $K_t$ (upside-down triangles) for wave frequency $f/f_0$ ranging from 2.2 to 4.4 and $L/\lambda$ from 0.3 to 1.1. Transmission decreases until reaching a minimum that coincides with a maximum of reflection at $f/f_0=3.0$. At higher frequency the transmission coefficient increases until reaching a maximum and there is little to no reflection. (b) Sum of the reflection and transmission coefficients $K_t+K_r$ (black diamonds). The plate dissipates between $20\%$ and $45\%$ of the incoming waves energy. (c) Plate deflexion $\xi$ as a function of the position $x$ for $f/f_0=$ 2.2; 2.9; 4.4.}
\label{Fig: Kt_Kr_plate}
\end{figure}

Figure \ref{Fig: Kt_Kr_plate} (a) shows the average, over five identical experiments, of the transmission and reflection coefficients, $K_t$ (upside-down triangles) and $K_r$ (triangles), respectively, as a function of wave frequency ranging from $f/f_0=2.2$ to $f/f_0=4.4$, corresponding to $L/\lambda$ ratio ranging from 0.3 to 1.1. Figure \ref{Fig: Kt_Kr_plate} (b) shows the sum of the coefficients $K_t+K_r$. In each of our experiments, the incident wave amplitude is close to $A=4$ mm. When waves interact with the plate, significantly less energy is transmitted after the obstacle. The values of the transmission coefficient range from $K_t\approx0.7$ to a minimum $K_t\approx0.3$ at around three times the resonance frequency of the plate. As the frequencies get higher, the transmission coefficient increases until it reaches $K_t\approx0.8$. The reflection coefficients $K_r$ mirror the tendency of $K_t$, and its maximum $K_r\approx0.3$ corresponds to $K_t$ at the same frequency. At higher frequencies there is little to no wave reflection. 

The sum $K_t + K_r$ as represented in Figure \ref{Fig: Kt_Kr_plate}(b) gives us information about the energy loss due to plate-waves interaction. Values are found to be between $K_t+K_r=0.8$ and $K_t+K_r=0.55$, corresponding to energy loss between $20\%$ and $45\%$ of the incoming waves energy. Thus, despite its simplicity, this system proves efficient to avoid waves transmission: in the best condition only $30\%$ of the waves energy is transmitted after the obstacle, and $\sim 40 \%$ of the energy is dissipated.

In the case of a rigid plate, under experimental conditions similar to ours, it has been shown that no reflection pattern were observable \citep{Polly2025}, highlighting that the plate motion, and therefore its flexibility, is a key element in reflecting water waves. They also showed that it is possible, in the case of a flexible plate, to attain $K_t + K_r$ values close to 0, thus configurations in which wave energy is totally dissipated. This corresponds to cases where the tip of the plate reaches the free surface and then breaks the waves with an effect similar to that of a beach. Our study also highlight the fact that plate flexibility enhances wave reflection, with $K_r$ values non zero and at best equal to 0.3. Figure \ref{Fig: Kt_Kr_plate}(c) shows experimental plate deflections $\xi$ at $f/f_0=$ 2.2; 2.9; 4.4. Maximum reflection occurs between the first and second modes. As the frequency increases, the plate's tip deflexion decreases. The motion of the plate is considerably reduced, leading to less interaction and $K_t+K_r$ values closer to one.

\subsection{Internal dissipation}

Wave energy dissipation is caused by two main processes. A part of the energy is dissipated internally, and an another part is externally dissipated due to the relative motion of the plate in water. We estimate the amount of energy dissipated by internal dissipation over one wave period, $\varepsilon_{in}$, corresponding to the amount of energy that is stored through the mechanical bending of the plate. It can be scaled as \citep{Nove2018, Polly2023_PhDthesis}:

\begin{equation}
    \varepsilon_{in} \sim 2 \rho_p e l \Gamma fL\xi^2 ,
\end{equation}
where $\Gamma$ is the internal damping coefficient and $\xi$ the plate deflection. Plate oscillations persist for around 15 seconds, leading to values of $\Gamma$ close to 0.1 s$^{-1}$, and plate deflection is at maximum $\xi=0.5$ cm. 

Since total wave energy density per unit area is $E_w=\frac{1}{2}\rho g A^2$, and since wave energy travels at group speed $c_g$, for a tank of width $b$ wave power $P_w$ can be expressed as:
\begin{equation}
    P_w=\frac{1}{2}\rho g b c_g A^2 ,
\end{equation}
and thus the energy of the incoming wave is:
\begin{equation}
    \varepsilon_{w} = P_w  T=\frac{1}{2f}\rho g b c_g A^2.
\end{equation}
If we compare $\varepsilon_{in}$ and $\varepsilon_{w}$ we obtain, at best:
\begin{equation}
    \frac{\varepsilon_{in}}{\varepsilon_{w}} = \frac{\rho_p}{\rho} \frac{4lLe\Gamma f^2 \xi^2}{gbc_gA^2} \sim 10^{-3}.
\end{equation}
The bending energy dissipated by the elastic plate is negligible compared to the dissipation observed from our $K_t+K_r$ results. Consequently, internal damping is not the main phenomenon at the origin of the observed dissipation. 

\subsection{Jet characteristics}

\begin{figure}[t]
\centering
\includegraphics[width=\linewidth]{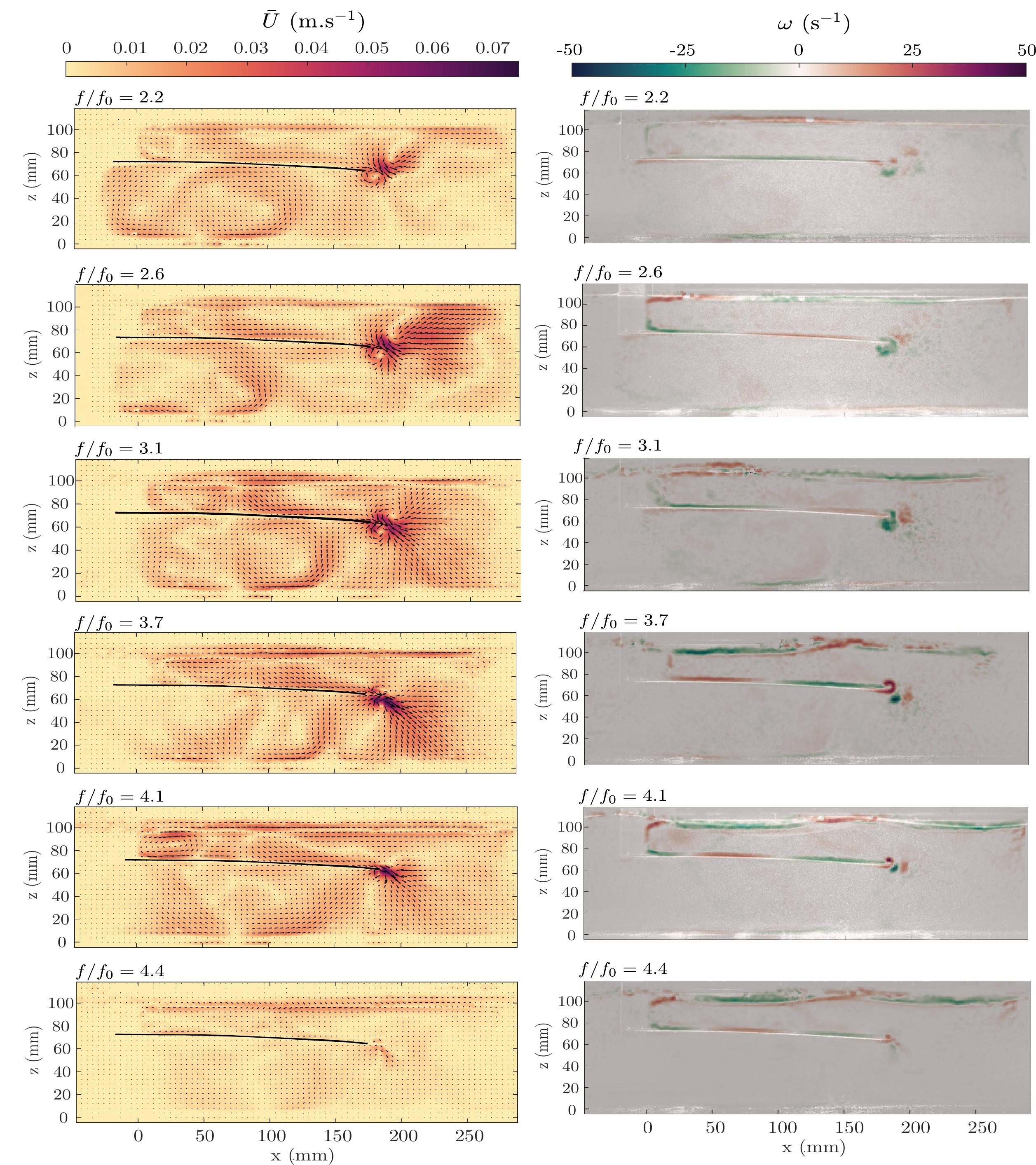}
\caption{On the left: Mean velocity fields $\bar U$ over 20 wave periods for $f/f_0=2.2$, $f/f_0=2.6$, $f/f_0=3.1$, $f/f_0=3.7$, $f/f_0=4.1$, and $f/f_0=4.4$. Arrows indicate mean fluid direction based on mean velocity along x and z axis. On the right: the corresponding vorticity fields at t/T = 1.}
\label{U et vort}
\end{figure}

PIV experiments are performed, using the method explained in Section \ref{section PIV}, in the same wave amplitude and frequency range conditions as before. Figure \ref{U et vort} shows mean velocity fields and vorticity fields at phase $\frac{t}{T}=1$, phase-averaged over 20 wave periods, for six different frequency cases $f/f_0=2.2$, $f/f_0=2.6$, $f/f_0=3.1$, $f/f_0=3.7$, $f/f_0=4.1$, and $f/f_0=4.4$. Arrows indicate the mean fluid direction based on mean velocity along $x$ and $z$ axes. The mean flow velocity norm $\overline{\rm U}$ is obtained by averaging over 800 instantaneous fields, which corresponds to 20 wave periods $T$:
 \begin{equation}
     \overline{\rm U}= \frac{1}{T} \int\sqrt{U_x^2+U_z^2} \mathrm{d}t .
 \end{equation}

For each case, we see alternate vortices at the tip of the plate and that the total mean flow is non-zero, especially where vortices appear. It is clearly the most intense at the tip of the plate corresponding to a jet formation, but also non zero above and below the plate with the apparition of cells.

 The intensity and orientation of the jet depend on the wave frequency. At $f/f_0=2.2$ and $f/f_0=2.6$ it is going downstream towards the free surface. At $f/f_0=3.1$ we see a very large downstream jet taking almost all the water depth, going horizontally but also towards the top and the bottom of the tank. At $f/f_0=3.7$ the jet is going downstream towards the bottom of the tank. At $f/f_0=4.1$ the jet is also going towards the bottom of the tank but with less intensity. Finally, at $f/f_0=4.4$ the jet has decreased visibly. In most cases, we observe the formation of small recirculation cells under the plate. For all frequencies, vortices remain close to the tip of the plate and are not advected, as is usually the case in rigid plate experiments such as those of Poupardin et al. \cite{Poupardin2012}.

The appearance of a mean flow in the case of a stationary rigid plate in a wave field has been numerically predicted by Carter et al. and Pinon et al. \cite{Carter2006, Pinon2017}, and experimentally observed using PIV by Poupardin et al. \cite{Poupardin2012}, where its formation has been attributed to vortex generation at the tip of the plate. In the case of an elastic plate forced not by waves but by a free stream flow, the generation of a wake is also observed, like in the study of D'adamo et al. \cite{D'adamo2022}. However, contrary to our study, in both rigid and elastic cases, no jets created at the end tip were reported going upward towards the free surface but instead always plunging towards the bottom. The interaction between the moving plate tip and the waves is probably causing this effect. 

\begin{figure}[t]
    \centering
    \includegraphics[width=\textwidth]{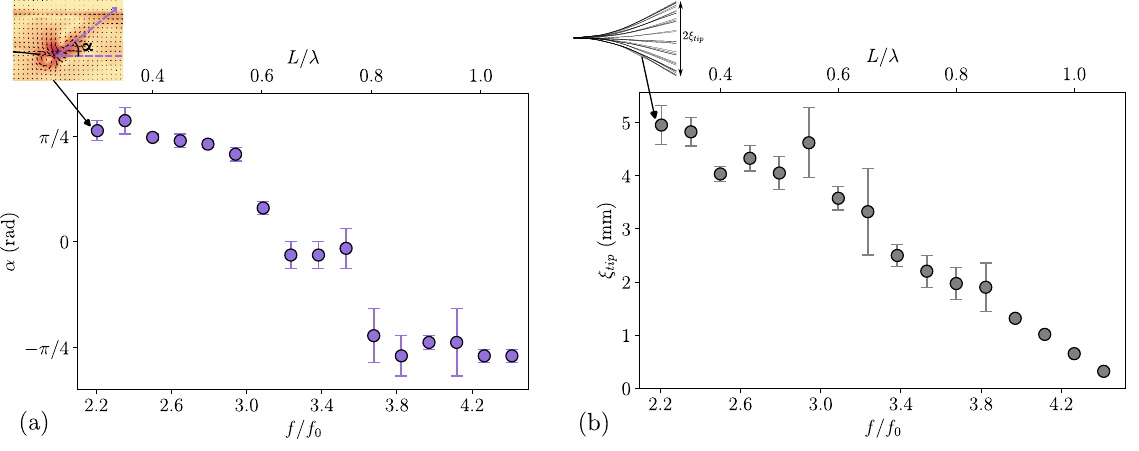}
    \caption{ (a): Angle $\alpha$ between the horizontal and the jet direction as a function of waves frequency, averaged from 2 different PIV experiments. Positive values correspond to a jet going upward and negative values to a jet directed downward. (b): Plate amplitude at the tip $\xi_{tip}$ corresponding to maximal deflection as a function of waves frequency, averaged from 2 different PIV experiments. For both (a) and (b) an example of measurement is represented for the case $f/f_0=$ 2.2.}
    \label{Fig:orientation}
\end{figure}

To characterise the jet direction depending on the wave frequency, the angle $\alpha$ between the horizontal and the jet direction has been represented in Figure \ref{Fig:orientation} (a). In the lowest frequency region the jet is directed upward, then shifts direction past $f/f_0=3.0$, and then is directed downward at higher frequencies. 

Figure \ref{Fig:orientation} (b) shows the plate deflection as a function of $f/f_0$. It can be observed that as the wave frequency increases the plate tip amplitude decreases, as we go further from the first bending mode of the plate. The smaller deflection observed at high frequencies could explain why the jet is directed downward, as the behaviour increasingly resembles that of a rigid plate. 

\subsection{Wave energy dissipation}\label{section Jet power}

\begin{figure}[t]
    \centering
    \includegraphics[width=\textwidth]{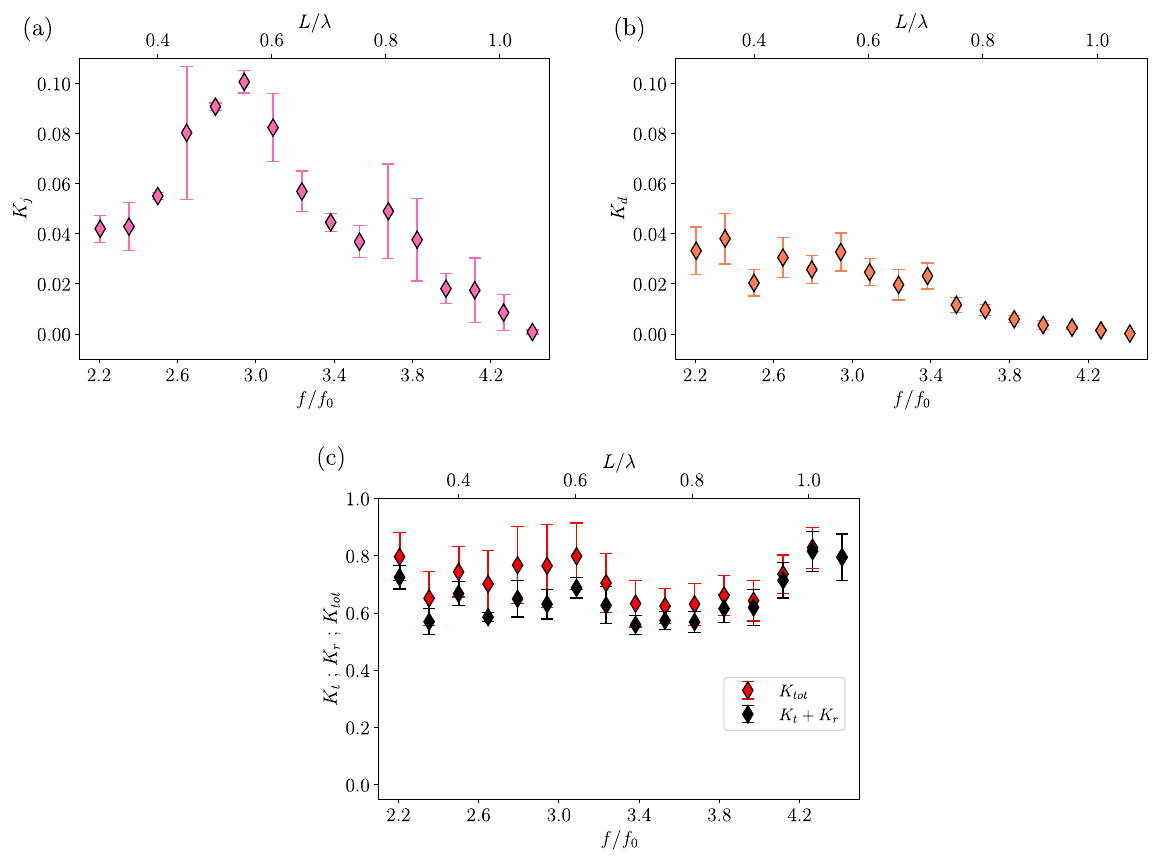}
    \caption{ (a): Ratio $K_j$ (pink diamonds) between jet power, $P_j$, and incoming wave power, $P_w$ as a function of wave frequency, averaged from 2 different PIV experiments. (b): Ratio $K_d$ (orange diamonds) between the energy dissipated by drag,  and incoming wave energy, $\varepsilon_w$ as a function of wave frequency, averaged from 2 different PIV experiments. (c): Sum $K_{tot}$ (red diamonds) of the reflection and transmission coefficients $K_r$ and $K_t$, the ratio from the fraction of energy inside the jet $K_j$, and the ratio from the drag dissipation $K_d$, compared to $K_t+K_r$ alone (black diamonds).}
    \label{Fig: Kj_Kd_Ktot}
\end{figure}

The jet shown by the PIV experiments could be the best candidate to explain the energy loss seen in Figure \ref{Fig: Kt_Kr_plate} (b). To measure the fraction of wave energy transferred to the jet, its power, $P_j$, can be compared to the incoming wave power, $P_w$. Energy in the jet is in the form of kinetic energy, consequently, $P_j$ can be estimated as the kinetic energy flux across a surface in the jet. It leads to writing:
\begin{equation}
    P_j = \frac{1}{2}\rho \int_{S} \bar{U}^2\, \bar{\mathbf{U}} \cdot \mathrm{d}\mathbf{S} ,
\end{equation}
 with $\mathbf{S}$ being a chosen orientated surface. In our case, $\mathbf{S}$ is a rectangular surface of width equals to the one of the tank $b$ and orientated normal to the jet direction $\alpha$ (Figure \ref{Fig:orientation} (a)). In doing so, the jet is assumed to be homogeneous in the $y$ direction. Under this assumption $P_j$ expression can be simplified as follows:
\begin{equation}\label{eq Pj}
    P_j = \frac{1}{2}\rho\, b \int_{-l_I/2}^{l_I/2} \bar{U}^2\, U_n \,\mathrm{d}l_I ,
\end{equation}
with $U_n$ the velocity component normal to the integration line $l_I$. $l_I$ is normal to the jet direction, and located inside the region corresponding to $z$ going from the bottom of the tank until 2 cm before the free surface, to only get the jet contribution and not free surface waves, and for $x$ between the tip of the plate to the end of the PIV domain. Let us define the dissipation coefficient as:
\begin{equation}
    K_j=\frac{P_j}{P_w}. \label{eq k} 
\end{equation}

Figure \ref{Fig: Kj_Kd_Ktot} (a) shows the dissipation coefficient $K_j$ with $P_j$ calculated using equation \ref{eq Pj} at the position where $P_j$ is maximal. $K_j$ values increase until reaching a maximum $K_j\approx 0.1$ around $f/f_0=3.0$ and $L/\lambda\sim0.5$, which also corresponds to the normalised frequency and wavelength values for which we observed a maximum reflection and minimum transmission in Figure \ref{Fig: Kt_Kr_plate}. $K_j$ then decreases to 0, which corresponds to the fact that at the highest frequencies the plate barely oscillates, thus resulting in the creation of a small jet with little energy, as seen in Figure \ref{U et vort}. In the region where the jet is the most intense we found that the plate dissipates around $40 \%$ of the wave energy, see Figure \ref{Fig: Kt_Kr_plate} (b), which is significantly more than $K_j=10 \%$ the maximum dissipation found in the jet.

We can also estimate the externally dissipated energy as a result of the relative motion of the plate in water, which corresponds to the drag dissipation arising from vortex generation. 

For a submerged elastic plate clamped at one edge, following the Morison equations \cite{morison_force_1950}, the drag power dissipation per unit surface area $P_d(x,t)$ is given by:
\begin{equation}
    P_d(x,t)=-\frac{1}{2} C_d \rho |U_{rel}(x,t)| U^2_{rel}(x,t) ,
\end{equation}
where $C_d$ is the drag coefficient and $U_{rel}$ is the relative velocity between the plate and the oscillatory flow. Assuming that the relative velocity is dominated by the plate motion, we have the following:
\begin{equation}
    U_{rel} = \dot{\xi}(x,t).
\end{equation}
Because of wave forcing, the plate motion is oscillatory and can be written as
\begin{equation}
    \xi(x,t)=\underline{\xi}(x) \cos(\omega t),
\end{equation}
therefore,
\begin{equation}
     P_d(x,t) = -\frac{1}{2} C_d \rho \omega^3 \left|\underline{\xi}(x) \sin(\omega t)\right|^3.
\end{equation}
By integrating this expression, we can obtain $\varepsilon_d$, the energy dissipated by drag:
\begin{equation}
\begin{split}
    \varepsilon_d &=  \int_{0}^{L} \int_{0}^{l} \int_{0}^{T} P_d(x,t) \,\mathrm{d}x \mathrm{d}y \mathrm{d}t \\
    &= - \frac{4}{3} \omega^2 C_d \rho l \int_{0}^{L} \underline{\xi}^3(x)\,\mathrm{d}x .
\end{split}
\end{equation}

In the case of vertical plates the drag coefficient $C_d$ can be estimated using the Keulegan-Carpenter number $KC$ \cite{graham_forces_1980} and \cite{merigaudModellingFarfieldEffect2024}: 
\begin{equation}
    C_d \propto KC^{-1/3} .
\end{equation}
For an oscillatory motion of amplitude $\xi$, the Keulegan-Carpenter number can be described as $KC=\omega \xi T/l=2\pi \xi/l$.
We can define an estimation of the drag dissipation coefficient as:
\begin{equation}
    K_d = \frac{\varepsilon_d}{\varepsilon_w} .
\end{equation}
Figure \ref{Fig: Kj_Kd_Ktot} (b) shows the drag dissipation coefficient $K_d$, averaged over two PIV experiments. The coefficient is found to be at best equal to $0.04$, but no clear maximum is found at $f/f_0$ like in the case of $K_j$. As frequencies increase, $K_d$ slowly decreases until it reaches $0$, coinciding with the decrease in plate deflection, see Figure \ref{Fig:orientation} (b).

Let us define the coefficient $K_{tot}$ corresponding to the sum of the wave transmission, reflection, the estimated energy from the jet, and drag dissipation:
\begin{equation}
    K_{tot}=K_t+K_r+K_j+K_d.
\end{equation}
This sum is presented in Figure \ref{Fig: Kj_Kd_Ktot} (c), as well as the sum $K_t+K_r$ as comparison. The two sources of external dissipation that we considered appear to account for some of the energy loss at intermediate and lower frequencies; however, it is clear that the estimates carried out for $K_j$ and $K_d$ are not sufficient to close the energy balance, as $1-K_{tot} \neq 0 $. One possible explanation for this difference comes from the assumption that the jet is homogeneous in the transverse direction $y$.

 \subsection{Transversal PIV experiments}



 \begin{figure}[t]
    \centering
    \includegraphics[width=0.9\textwidth]{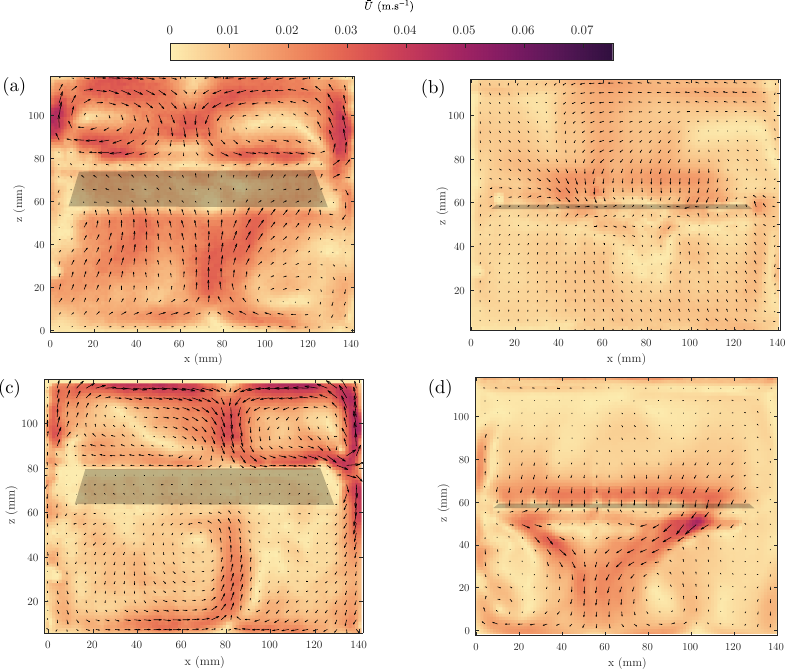}
    \caption[Transversal mean velocity fields.]{On the left: Mean velocity fields $\bar U$ over 20 wave periods for $f/f_0=2.2$ (a), $f/f_0=2.9$ (c) in the $y-z$ plan at the middle of the plate. On the right: Mean velocity fields $\bar U$ over 20 wave periods for $f/f_0=2.2$ (b), $f/f_0=2.9$ (d) in the $y-z$ plan 1 cm after the tip of the plate. The plate is represented in light gray.}
    \label{Fig: transverse}
\end{figure}

To get a better idea of the vortex dynamics transverse to the direction of wave propagation, additional PIV experiments were conducted in the $y-z$ plane. Figure \ref{Fig: transverse} shows the mean velocity fields for $f/f_0=2.2$ and $f/f_0=2.9$ either in the $y-z$ plane at the middle of the plate for Figures \ref{Fig: transverse} (a) and (c) or 1 cm after the end of the plate for Figures \ref{Fig: transverse} (b) and (d). Motion in the $y - z$ plane appears quite clearly, as recirculation cells can be observed above and below the plate. The velocities are significant, close to $3$ cm.s$^{-1}$ in the plane at the middle of the plate. Vortices also forms on the lateral edges of the plate, in the space between the object and the walls of the tank, with a velocity going up to $5$ cm.s$^{-1}$. 
 
Other studies, such as Raspa et al. \cite{Raspa2014}, have also investigated the appearance of complex flows at the edges by performing PIV experiments on a self-propelled elastic foil that freely swims in a still fluid. They show that the major part of the total drag opposing the motion of the foil comes from trailing vortices that roll up at the lateral edges, and that this vortex-induced drag depends on the foil's aspect ratio. In our case, this effect, in addition to the relative motion caused by the wave flow, might be responsible for the rest of dissipation of wave energy measured in Section \ref{section Jet power}.

\section{Discussion and conclusion}

This study characterised the interaction between a submerged flexible plate clamped at one edge and water waves. Experiments proved that such a configuration efficiently dissipates wave energy, supplementing other work showing how flexibility enhances dissipation compared to more rigid structures \citep{Polly2025,Kumar2007}. PIV experiments have been carried out to understand the physical process behind the wave dissipation. We showed that a vortex pair is periodically shedded at the tip of the plate and generates a strong jet whose direction depends on wave frequency. In our lowest frequency range we observed the jet going upward, before shifting orientation and going downward at higher frequencies. In the case of a rigid plate, some studies highlighted the creation of a downstream vortex pair leading to a strong jet directed towards the flume bottom, as well as the presence of recirculation cells under the plate \citep{Poupardin2012, Pinon2017}. Such configurations create stagnation points that could lead to scouring or accumulation in the case of a sedimentary bed. In the case of our flexible plate, we also observe the formation of recirculation cells under the plate, but stagnation points in the region towards the plate's end can be avoided when the jet is going upward. This configuration, which dissipates a significant amount of energy, could be of particular interest for WEC creation aiming to minimise the impact on marine life which is sensitive to scouring phenomena, as well as minimising flows towards the structure foundations. 


\bibliography{biblio_plate_waves}

\end{document}